\documentclass[aps,prx,twocolumn,noeprint,floatfix,superscriptaddress]{revtex4-2}

\usepackage{amsmath}
\usepackage{amssymb}
\usepackage{graphicx}
\usepackage{hyperref}
\usepackage{color}
\newcommand{\figref}[1]{Fig.~\ref{#1}}

\renewcommand{\eqref}[1]{Eq.~(\ref{#1})}

\newcommand{\llangle}{\langle\langle}
\newcommand{\rrangle}{\rangle\rangle}
\newcommand{\bvec}[1]{\mathbf{#1}}

\begin{document}

\title{Engineering Topological Phases Guided by Statistical and Machine Learning Methods}
\author{Thomas Mertz}
\author{Roser Valent\'{i}}
\affiliation{Institut f\"ur Theoretische Physik, Goethe-Universit\"at Frankfurt, Max-von-Laue-Stra{\ss}e 1, 60438 Frankfurt am Main, Germany}

\begin{abstract}
The search for materials with topological properties is an ongoing effort. In this article we propose a systematic statistical method supported by machine learning techniques that is capable of constructing topological models for a generic lattice without prior knowledge of the phase diagram. By sampling tight-binding parameter vectors from a random distribution we obtain data sets that we label with the corresponding topological index. This labeled data is then analyzed to extract those parameters most relevant for the topological classification and to find their most likely values. We find that the marginal distributions of the parameters already define a topological model. Additional information is hidden in correlations between parameters. Here we present as a proof of concept the prediction of the Haldane model as the prototypical topological insulator for the honeycomb lattice in Altland-Zirnbauer (AZ) class A. The algorithm is straightforwardly applicable to any other AZ class or lattice and could be generalized to interacting systems.
\end{abstract}

\maketitle

\section{Introduction}
In recent years machine learning techniques have enjoyed growing attention
among the physics community. Fueled by popular success in automation across a
wide variety of industrial applications,
implementations to fundamental research have been
proposed. Apart from, for instance,
the popularized computer vision application in black hole
research \cite{EHT2019}, a lot of effort has been devoted to increase the
efficiency of available algorithms, such as Monte Carlo \cite{Huang2017,Xu2017,Broecker2017,Pilati2019,Song2020} or
Density Functional Theory \cite{Jinnouchi2019a,Jinnouchi2019b,Nagai2020,Denner2020}. Moreover, the concept of machine
learning has been shown to be able to grasp even the very complex nature of
topological phases, finding the correct order parameter by itself
\cite{Wang2016,Carrasquilla2017,Kenta2020}. Successful reports of both, supervised and unsupervised
paradigms have been published recently \cite{Deng2017,Nieuwenburg2017,Zhang2018,Sun2018,Lian2019,Rodriguez-Nieva2019,Rem2019,Balabanov2020,Greplova2020,Scheurer2020,Che2020,Boesch2020}. An overview in terms of an extensive review of machine learning applications to condensed matter physics is also available \cite{Carrasquilla2020}.

In this work, we are proposing a different scheme where we lay emphasis on
minimal bias. Rather than speeding up a (in this case) manageable computational
task, we aim at machine-assisted learning of previously unknown information
using the toolkit of data science/statistics. 
Specifically we construct, following this scheme, topological models for honeycomb lattices. 
Dissecting first the well-known Haldane model \cite{Haldane1988} to benchmark and
validate our findings, we then look at the most general
model on a honeycomb lattice and use our analysis to extract a topological
prototype model for each individual class label. These generated models turn
out to be exactly of the Haldane type. This procedure can be generalized
to any generic lattice and shows that topological models can be ``learned" from the statistics of a randomized data set, not only by a machine since the result is readily comprehensible.

The paper is organized as follows. In Section \ref{sec:data_generation} we discuss the generation of our data and features. Section \ref{sec:statistical_method} contains the motivation and definition of the quantities used to extract information from the data, which is then applied to the Haldane model in Section \ref{sec:haldane_model} and a general honeycomb lattice in Section \ref{sec:honeycomb}.

\section{Data generation}
\label{sec:data_generation}
We first start by introducing some definitions of quantities
that will be used throughout the paper.
We define ``data" as a set of feature vectors $\bvec{x}_i$ with dimension $n_f$ (number of features), which can be stacked into a data matrix $X = (\bvec{x}_0, \bvec{x}_1, \bvec{x}_2, \ldots)^T$ with dimensions $n_s \times n_f$, where $n_s$ is the number of samples or data points. The corresponding labels are stored in variables $y_i\in \mathbb{Z}$, which can be written as a single vector $Y$. We denote a specific feature as $x_j:= X_{ij}=[\bvec{x}_i]_j$, where we omit the sample index if possible.
The feature matrix $X$ and the label vector $Y$ are related by a non-linear transformation $f$, such that $f(X) = Y$.

Here, we compute the label from $X$ by calculating the topological index (in this case the Chern number) from the model specified by $\bvec{x}_i$ (the $i$-th row of $X$)
\begin{equation}
	y_i = C(H_k(\bvec{x}_i)),
\end{equation}
where $H_k(\bvec{x}_i)$ is the Bloch Hamiltonian of the model and $f=C\circ H$. 
The label $y_i$ serves as a classifier that allows us to separate the data into different sets. We will then analyze the differences between these different data sets by statistical means without further reference to the label.

Data points are generated by choosing a reference point $x_\mathrm{ref}$ and subsequently sampling perturbations $\delta_i$ to this point from suitable random distributions to create a cloud of data points around $x_\mathrm{ref}$. For each point we store both $x_i=\delta_i$ and the label $y_i$.

\subsection*{Choice of features}
A model describing a quantum material
is typically represented in terms of tight-binding parameters, where symmetries are
already accounted for.
A general representation applicable to multiorbital materials is that of hopping matrix
elements or overlap integrals of orbitals. By denoting every parameter
$t_{ij}(R)$ with the displacement vector $R$ between the different orbitals,
in addition to the site-orbital indices $i,j$, we have more parameters at our
disposal which allow us to break symmetries and potentially discover unknown
topological phases. 
Our feature vector thus consists of all $t_{ij}(R)$ up to a cutoff distance $|R|$. We note that this choice would pose a great
challenge to typical machine learning applications, since not only the computation of the Chern number, but also the diagonalization and 
construction of the Hamiltonian has to be learned, which would require an extremely complex model. By choosing this most
general data set (model parameters, topological class label) we make sure that we can learn about the relation of the
topological classification to the physical parameters of the system. In contrast to a similar approach, where machine learning was used to speed up the construction of a 
tight binding model \cite{Peano2019}, we are here only interested in extracting previously unknown information from the data that is not otherwise attainable.

We note that, concerning our study on topological
phases,
this description of quantum materials encloses both,
non-interacting electron systems as well as interacting electron
systems where the concept of topological Hamiltonian is applicable \cite{Wang2012,Mertz2019}. Since the validity of this topological Hamiltonian is restricted to the weak to intermediate regime of correlations, the self-energy is not strongly momentum-dependent \cite{Mertz2018}. The weak sensitivity of the topological invariants w.r.t.~this momentum-dependence \cite{Mertz2019} suggests that modifications of the local hopping parameters ($R=0$) can also describe correlation effects.

For simplicity we work with real features $\bvec{x} \in \mathbb{R}^{n_f}$.
However, overlap integrals $t_{ij}$ are generally complex numbers, not
necessarily real, therefore 
we impose a mapping $g: \mathbb{C} \rightarrow \mathbb{R}^2$
to obtain a real feature vector. For complex parameters natural choices are
either $(\mathrm{Re}(x_i), \mathrm{Im}(x_i))$ or $(|x_i|, -i\log(x_i/|x_i|))$.
Since we don't know a priori
which is the better choice, we will use in what follows
both mappings. For
strictly real features we just take the real part of the definition above.

\begin{figure}[h]
	\includegraphics[scale=1]{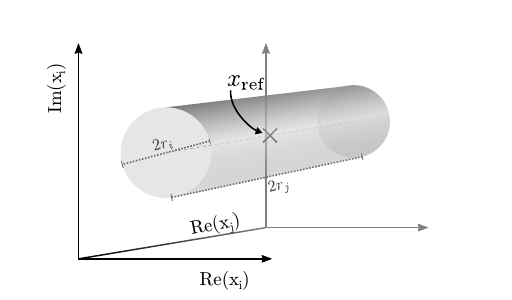}
	\caption{Features are uniformly distributed over a circular region with radius $r_i=\alpha|x_\mathrm{ref}^i|$ around the reference point $x_\mathrm{ref}$. The spread in the real parameter $x_j$ is given by $r_j=\alpha|x_\mathrm{ref}^j|$.}
	\label{fig:distribution}
\end{figure}

In order to be as unbiased as possible we choose a uniform probability
distribution for sampling our features. However, since we do not want to
generate too many extremely unphysical data points, we set the sample space
independently for each feature $x_i$ as $\Omega_\alpha =
B_{\alpha|x_\mathrm{ref}^i|}(x^i_\mathrm{ref}) \subset \mathbb{C}$, where
$B_r(x)$ denotes the solid sphere with radius $r$, centered at $x$. The external
parameter $\alpha := r_i/x^i_\mathrm{ref}$ is the ratio between the spread of the
data and the initial value, cf.~\figref{fig:distribution}.
The probability
density function (PDF) is then given by the uniform distribution on the sample space $\Omega_\alpha$ \begin{equation}
	\rho_\alpha(x) = U(\Omega_\alpha).
	\label{eq:prob_density}
\end{equation}
This choice guarantees our two requirements, namely being unbiased and,
preserving at least some amount of physicality of our model given a proper choice of the reference point $x_\mathrm{ref}$. The term ``physicality" here refers to closeness to a known physically reasonable configuration, that for example corresponds to a material. If we sampled instead over arbitrary domains of values we would take into account only more of those data points that do not conform with a tight-binding representation (i.e.~long-ranged hoppings much larger than short-ranged).

\section{Statistical method}
\label{sec:statistical_method}
After generating a reasonably large data set,
we proceed with the analysis of the information contained within.

In the first step we extract the most
characteristic features from the labeled data. We can define the relevance of
a feature through the discrimination between different labels. 
Restricting the data set to a specific
class label will reduce the entropy of certain features, which becomes clear
if we interpret the feature data and the label data as separate random
variables $X$ and $Y$, respectively $H(X | Y) = H(X) - I(X; Y)$.  One expects
the reduction in entropy, given by the mutual information $I(X;Y)$
(Eq.~\ref{eq:mutual}), to be a
measure for the importance of a feature. Given our particular data at least, we find that this definition lacks robustness with respect to noise and is therefore inapplicable to a general case.
We can nevertheless inspect the probability
distributions, or rather the frequency or empirical probability, of the
individual features. 

\begin{figure}[h]
	\includegraphics[scale=1]{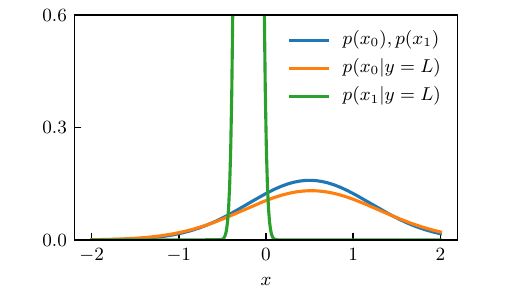}
	\caption{Illustration of a probability distribution function for two features $x_0$ and $x_1$. When restricted to the data subset with class label $L$ the distribution of feature $x_{1}$ deviates significantly from the base distribution, i.e.~the feature is more important for the classification.}
	\label{fig:illustration_pdf_comparison}
\end{figure}

We restrict our discussion to weakly correlated features and comment on possible
treatment of correlations beyond that further below. Comparing
probability distributions between different classes should thus yield a
measure of importance for the individual features. An illustration of this
motivation is provided in \figref{fig:illustration_pdf_comparison}, where we
show the difference between less important features ($x_0$) and important
features ($x_1$). The projection onto the subspace corresponding to label $L$
results in only a minor modification for the former, while the latter deviates
substantially.  

We quantify the difference between two probability distribution
 functions  $p(x), q(x): \mathbb{R}\rightarrow [0, 1]$ in terms of the Bhattacharyya distance \cite{Bhattacharyya1943}
\begin{equation}
	D_B(p, q) = -\log\left[\int\limits_{-\infty}^{\infty} \sqrt{p(x)q(x)} ~\mathrm{d}x\right],
	\label{eq:bdist}
\end{equation}
which satisfies $D_B(p, q)\geq 0$ and $D_B(p, q)=0$ iff $p=q$. Thus, according to the argument above, larger values of $D_B$ represent a larger importance of the feature. This measure has several advantages over the use of divergences in signal selection \cite{Kailath1967} and is also used for feature extraction for image recognition \cite{Choi2003,Reyes2006}. We note that, mathematically speaking, $D_B$ is not a distance since it does not satisfy the triangle inequality. The related Hellinger distance $D_H(p,q)=\sqrt{1-e^{-D_B}}$ is a true distance function. In our calculations, though, the Bhattacharyya distance proved to be more effective.

By only considering those features with the highest importances we can perform a dimensional reduction on the data set. One could now introduce new features that have an e.g.~polynomial dependence on the original features ($x_{i0}, x_{i1}, ..., x_{iN}, x_{i0}x_{i1}, ...$). This can be repeated to find a more optimal representation of the data. Albeit conceptionally simple, an actual implementation is not straightforward, though feasible since all operations required in a single step are basically $O(N)$.

\begin{figure}[h]
	\includegraphics[scale=1]{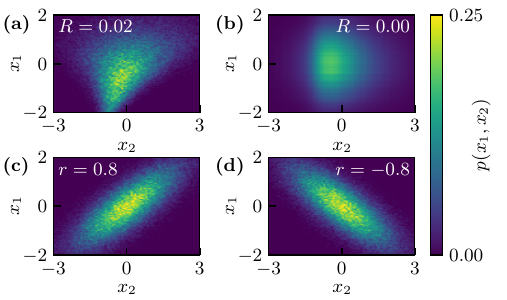}
	\caption{\label{fig:correlations_illustration}Illustration of the redundancy $R$ [\eqref{eq:redundancy}] (top row) and the Pearson correlation coefficient $r$ [\eqref{eq:pearson}] (bottom row). In \textbf{(a)} a joint probability density function for two dependent random variables is shown. The redundancy is nonzero. The product of the corresponding marginal distributions is shown in \textbf{(b)} and clearly differs from the true joint distribution. The redundancy between independent variables vanishes. \textbf{(c)} and \textbf{(d)} are examples for joint distribution functions for positively and negatively correlated variables. Note the respective sign of the PCC.}
\end{figure}

Without introducing the aforementioned features it is unclear how this approach performs if features are correlated, i.e.~if phase separation lines do not lie along parameter axes. We employ a twofold analysis based on the statistical dependence and correlation, which indicate relations between different random variables. In \figref{fig:correlations_illustration}(a,b) we illustrate for example that statistical dependence (a) means that the distribution function for one parameter depends on that of the other, whereas this is not the case for independent parameters (b). Correlations on the other hand specify a particular nature of statistical dependence as seen in \figref{fig:correlations_illustration}(c,d). Here, we measure the statistical dependence in terms of a normalized variant of the mutual information, that we call \emph{redundancy}
\begin{equation}
	R = \frac{I(X; Y)}{H(X, Y)},\quad R\in [0, 1],
	\label{eq:redundancy}
\end{equation}
where $I$ is the mutual information
\begin{equation}
	I(X;Y) = \int\limits_{-\infty}^{\infty} p(x, y) \log\left[\frac{p(x, y)}{p(x)p(y)}\right]\mathrm{d}x\mathrm{d}y.
	\label{eq:mutual}
\end{equation}
and $H(X,Y)$ the joint entropy of random variables $X, Y$
\begin{equation}
	H(X,Y) = \int\limits_{-\infty}^{\infty} p(x,y) \log\left[p(x,y)\right]\mathrm{d}x.
\end{equation}
Alternatively, when features are dependent on one another we quantify the nature of correlations in terms of the Pearson correlation coefficient (PCC)
\begin{equation}
	r_{X_i,X_j} = \frac{\mathrm{Cov}(X_i,X_j)}{\sqrt{\mathrm{Var}(X_i)\mathrm{Var}(X_j)}},
	\label{eq:pearson}
\end{equation}
which can differentiate uncorrelated and positively/negatively correlated features. Technically, the PCC is only good for a linear dependence, considering the limited window of parameter values, though, this method is still applicable and proves to be reliable enough.

While statistical independence and correlations are two different quantities, here we usually use the term ``correlations'' for both. This simplification is fine since we always look at statistical independence first and discuss statistical correlations only in case of dependent features.

We note that at this point we choose to simplify and only take into account correlations between pairs of features. Generalizations to higher order correlations exist, such as the total correlation \cite{Watanabe1960}, however, it is clear that the higher the order of the correlation function the more obvious the result will be in terms of a finite value, since a large number of random variables is less likely to be independent compared to a pair. At the same time the information content of such quantities decreases since one loses the fine granularity. Finding the right balance between complexity and information content is thus very difficult but necessary to fully understand the interplay between parameters.

\section{Benchmark case: Haldane model}
\label{sec:haldane_model}
\begin{figure}[h]
	\includegraphics[scale=1]{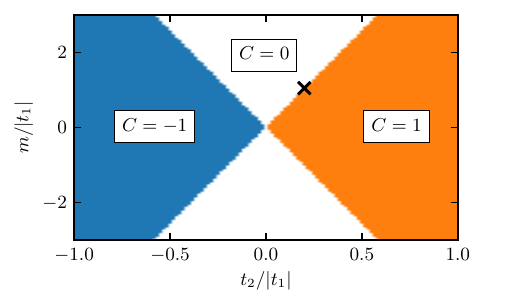}
	\caption{Phase diagram of the Haldane model for $\phi=\pi/2$ in terms of next-nearest neighbor hopping $t_2$ and mass $m$. Starting out from the trivial phase (0), one can reach a non-trivial phase by changing either $m$ or $t_2$ or both. The reference point $x_\mathrm{ref}$ is marked by $\mathbb{\times}$.}
	\label{fig:haldane_phase_diagram}
\end{figure}

The Haldane model \cite{Haldane1988} is defined as
\begin{align}
\begin{split}
	H &= t_1\sum_{\langle ij\rangle} c_i^\dagger c_j + t_2 \sum_{\llangle i,j \rrangle} e^{i\phi_{ij}} c_i^\dagger c_j \\
	&\quad+ m \sum_i \mathrm{sign}(i) c_i^\dagger c_i,
\end{split}
\label{eq:haldane_model}
\end{align}
where $\phi_{ij}=\pm 1$ for counterclockwise or clockwise hopping within a hexagon. This ensures a staggered flux pattern that results in a vanishing overall magnetic field. Since both time-reversal and particle hole symmetry are broken, \eqref{eq:haldane_model} is an example of a topological insulator in AZ class A \cite{Altland1997,Chiu2016}.
One obtains a rich phase diagram, see \figref{fig:haldane_phase_diagram} for $\phi=\pi/2$, with a trivial insulator ($C=0$) at $m/|t_2| > a$, a Chern insulator with topological index $C=+1$ at $0 < |m|/t_2 < a$ and a Chern insulator with topological index $C=-1$ at $a < |m|/t_2 < 0$. The value of $a\in\mathbb{R}$ depends on $\phi$ and will approach 0 when reaching $\phi=n\pi$ for $n\in \mathbb{Z}$.

Implicitly, \eqref{eq:haldane_model} assumes a perfect honeycomb. If we relax this requirement we obtain a model with 11 independent parameters
\begin{equation}
	H = \sum_{\langle i,j\rangle} t_{1}^{ij} c_j^\dagger c_i + \sum_{\langle\langle i,j\rangle\rangle} t_{2}^{ij} c_j^\dagger c_i + \sum_i \varepsilon_i c_i^\dagger c_i,
\end{equation}
namely three nearest neighbor terms $t_1$, six next-nearest-neighbor terms $t_2$ and two onsite terms $\varepsilon_i$ with $\varepsilon_A-\varepsilon_B=2m$. Due to the requirement that the Hamiltonian be hermitian, $\varepsilon_i$ must be real. 
All other parameters are sampled as complex values. Thus, we have nine complex and two real features or equivalently 20 real features. In order to fix the energy scale, one of the onsite terms should always be set to zero, which leaves a total of 19 real features.

The order of the complex features is defined in the following way
\begin{equation}
	\bvec{x}_i = (0, m, t_{1}^1, t_{1}^2, t_{1}^3, t_{2}^1, t_{2}^2, t_{2}^3, t_{2}^4, t_{2}^5, t_{2}^6),
\end{equation}
where the superscript index differentiates the three (six) different values of $t_1$ ($t_2$). The leading 0 corresponds to the onsite energy $\varepsilon_A$.
We first fix as a reference point the coordinates of the Haldane model with
$m/t_1=1.05$, $t_2/t_1=0.2$, which lies just barely inside the trivial phase
region, cf.~\figref{fig:haldane_phase_diagram}. In feature space this can be
written as $\mathrm{Re}(x_\mathrm{ref}) = (0, 1.05, 1, 1, 1, 0, 0, 0, 0, 0,
0)$ and $\mathrm{Im}(x_\mathrm{ref}) = (0, 0, 0, 0, 0.2, -0.2, 0.2, -0.2, -0.2, 0.2)$. The sign change of the next-nearest neighbor term is due to
Haldane's requirement that the total flux be zero.

We run a fully unbiased sweep, where we draw samples in this 19-dimensional
space from the uniform probability density function \eqref{eq:prob_density}
with $\alpha=2$, which, on the one hand,
is large enough to allow for a sign change, but, on the other hand,
is small enough not to require an unfeasible number of samples. For each sample
the Chern number is computed and stored in the label vector. By using a
binning analysis we extract the frequency of different values for all features
within the different class labels.

\begin{figure}
	\includegraphics[scale=1]{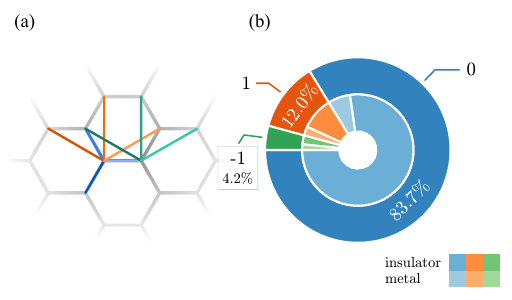}
	\caption{\textbf{(a)} Nearest- and next-nearest-neighbor
	hopping terms accounted for in the honeycomb lattice. We
	draw independent parameters in different colors. \textbf{(b)} Percentage of
	samples categorized by the topological class label (outer ring) and
	the corresponding fraction of insulators/metals (inner ring). Here, we
	find only $y=0, 1, -1$ in the surveyed region. The total sample size is
	$n_s = 10^7$.}
	\label{fig:haldane_piechart_unbiased}
\end{figure}
\begin{figure}
	\includegraphics[scale=1]{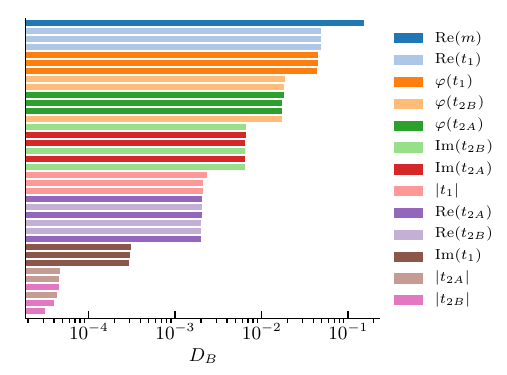}
	\caption{Importance scores in terms of the Bhattacharyya distance for all (real) features. Here, we take into account only the topological class with Chern index $C=1$. Most relevant are apparently the mass $m$, real part and phase of $t_1$ and the phase of $t_2$. We plot a separate bar for every individual hopping vector, equal colors indicate equal lengths.}
	\label{fig:haldane_importance_unbiased}
\end{figure}

We find a considerable number of non-trivial samples,
cf.~\figref{fig:haldane_piechart_unbiased}, even in our totally unbiased
approach. This number is large enough to extract useful statistical
information. With the given $x_\mathrm{ref}$ we obtain two topological phases
(1, -1), however, data with -1 is less abundant due to the larger distance of
$x_\mathrm{ref}$ from that phase region. The importance scores
[\eqref{eq:bdist}] computed from the distributions are shown in
\figref{fig:haldane_importance_unbiased}. Here, we show both mappings to the
real axis (Re/Im, $|.|$/$\varphi$). The mass $m$ is apparently most important,
following behind are Re($t_1$) and the phase of $t_2$, $\varphi(t_2)$. Since the imaginary part of $t_1$ ranks comparatively low the phase information must relate to the sign. Obviously the real part contains the information about the sign, so we choose here the real part.
Therefore, we can restrict the following discussion to the reduced set of 10
out of the total 39 features.
We have also trained a random forest classifier on the data and extracted importance scores via the permutation importance, cf.~e.g.~\cite{Guterding2020}, which resulted in a very similar ranking. The advantage of the present method is that we skip the costly training phase entirely.

\begin{figure}
	\includegraphics[scale=1]{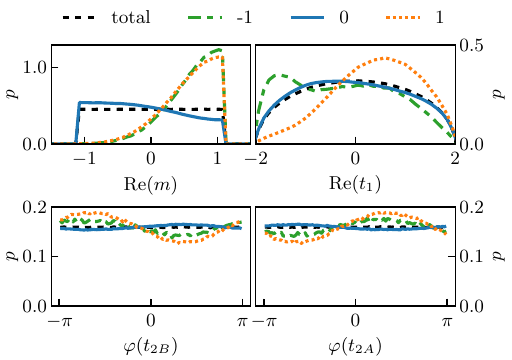}
	\caption{Relative frequency (approximate probability density function) for four important features. We chose here the mass $m$, the real part of a nearest-neighbor hopping $t_1$ and the phases of two next-nearest neighbor hoppings connecting $A$ and $B$ sites, respectively. For all terms we observe a clear distinction of the PDF of the non-trivial phase ($C=1, -1$) from that describing the trivial phase ($C=0$).}
	\label{fig:haldane_distribution_unbiased}
\end{figure}

Given the importance scores we inspect the underlying distributions more closely. These are expected to show a certain symmetry such that e.g.~nearest neighbors are interchangeable. While this is true, here, next-nearest neighbors are divided into two distinct groups, namely those that connect $A$ and $B$ sites, respectively. Thus, we end up with four distinct distributions, for which we show the measured values in \figref{fig:haldane_distribution_unbiased}.

Having extracted those features that show the clearest statistical response to the change of the topological label or vice versa, the question about the relationships between different features remains open. Due to the extremely unbiased approach and the large number of degrees of freedom therein it is clear that there will be no clearcut distinction between the different phases. To understand this we assume that the value of a feature can fall into separate intervals corresponding to the different phases. Since the number of features is large it is very likely that changing another feature moves the intervals around. Marginalizing over all other features then leaves us with a blurred out distribution that can no longer confidently distinguish phases. Therefore, we aim here at only finding the characteristic behavior. As a consequence of the large number of correlated features the correlations between any pair of features are rather small. This is interesting as it demonstrates the stability of the topological phase with respect to noise. Apparently, changing a single hopping parameter---even drastically---can leave the topological phase unchanged. This is also visible in the joint PDFs between any pair of features, which are all close to the independent PDF $p(x_i, x_j)=p(x_i)p(x_j)$, resulting in small redundancy values. Correlations between many (if not all) features should be present and the corresponding joint PDF contains the complete information about the classification. Nevertheless, the joint PDFs are extremely difficult to interpret.

Finding a prototype feature set for a specific label can intuitively be done by taking the mean of the corresponding data points in case of a symmetric distribution or the peaks in case of an asymmetric distribution. However, this does not always lead to a correct classification, since correlations are neglected.
Given the measured frequency of a particular set of features it is apparently more likely that for a single sample most values lie close to the respective peaks, while only few deviate significantly. Taking into account the correlation coefficient between the features we can distinguish between actual correlation and noise.

\begin{figure}
	\includegraphics[scale=1]{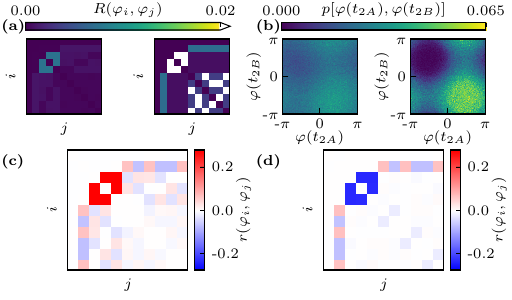}
	\caption{\textbf{(a)}~Redundancy $R$, \eqref{eq:redundancy}, shown here
	for the phases of all parameters for the unbiased (left) and biased (right)
	topological data set. The nearest neighbor hoppings show a small
	redundancy in the unbiased data. This is not the case for the
	next-nearest neighbor hoppings. Imposing the bias on the data reveals a
	redundancy between $t_{2A}$ and $t_{2B}$. \textbf{(b)}~Joint probability density function $p[\varphi(t_{2A}), \varphi(t_{2B})]$ for the
	next-nearest neighbor hoppings for unbiased (left) and biased (right)
	data sets. Apparently, noise due to the large
	number of degrees of freedom for the six next-nearest neighbor hoppings
	reduces the contrast in the PDF and therefore reduces the measured
	redundancy. Bottom: Pearson correlation coefficient
	[\eqref{eq:pearson}] for the \textbf{(c)} $C=1$ and \textbf{(d)} $C=-1$ 
	phase. We observe positive and negative correlations between the
	nearest-neighbor hoppings, respectively.}
	\label{fig:pearson_haldane_unbiased}
\end{figure}
We investigate the statistical dependence of the parameters in terms of
the redundancy (Eq.~\ref{eq:redundancy}) in \figref{fig:pearson_haldane_unbiased}(a), and the Pearson correlation coefficient (Eq.~\ref{eq:pearson}) in \figref{fig:pearson_haldane_unbiased}(c, d). In addition we illustrate the corresponding joint PDF between a pair of features in \figref{fig:pearson_haldane_unbiased}(b).
We find that the nearest-neighbor hoppings are positively correlated in the topological class $C=1$ [see \figref{fig:pearson_haldane_unbiased}(c)], which indicates that the three different values are similar. For the $C=-1$ class [see \figref{fig:pearson_haldane_unbiased}(d)], however, we find the opposite sign, i.e.~the hopping values are negatively correlated. This means that one or two values have the opposite sign w.r.t.~the mean.

Given this information we can construct effective models for the two classes $C=1$
and $C=-1$. To this end we reduce the complexity further by assuming a symmetry
between the $t_1$ and $t_2$ features. While this is not necessary, as shown by
the statistical independence of the parameters in the data [\figref{fig:pearson_haldane_unbiased}(a)], it greatly improves
the interpretability of the data. Depending on the topological class label and
the associated correlations, the hopping terms are either equal or have
opposite signs. The $t_2$ values are split into two independent groups based on
the distinct PDFs obtained in the unbiased run.  This reduced set of
parameters contains seven independent degrees of freedom vs the original 19.

\begin{figure}
	\includegraphics[scale=1]{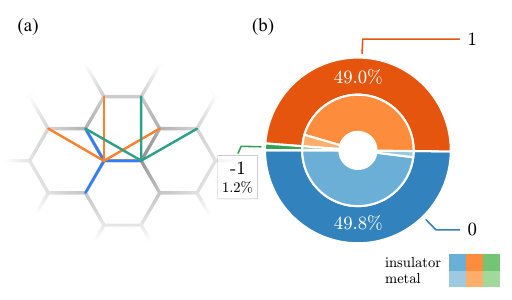}
	\caption{\textbf{(a)} Hopping parameters taken into account. Terms related by symmetry are colored equally. \textbf{(b)} Percentage of samples from the improved model categorized by the topological class label (outer ring) and the fraction of insulators/metals therein (inner ring). In the surveyed region we find almost exclusively insulators with labels $y=0, 1$; the number of $y=-1$ samples is statistically irrelevant. The total sample size is $10^6$.}
	\label{fig:haldane_piechart_biased}
\end{figure}

The improved model with reduced complexity is given by four distinct parameters, i.e.~one real onsite term, one complex nearest-neighbor term and two complex next-nearest-neighbor terms. Due to the reduced complexity, a good statistics is obtained at lower sample sizes, allowing for a quicker evaluation. In \figref{fig:haldane_piechart_biased} we show that the visibility of the non-trivial topological $C=+1$ phase in the data has greatly improved, which validates the choice of symmetries for our biased model.
We use the data obtained from this run to finally settle exemplary values for the prototype model.

\begin{figure}
	\includegraphics[scale=1]{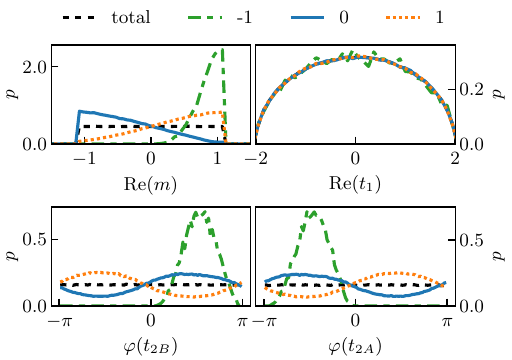}
	\caption{Relative frequency (approximate probability density function)
	for the mass $m$, a nearest-neighbor term $t_1$ and two next-nearest
	neighbor terms $t_2$ for the biased data. Compared to the unbiased data
	(see Fig.~\ref{fig:haldane_distribution_unbiased}) the nearest-neighbor term is suddenly completely
	indistinguishable between different phases, while the contrast of the
	next-nearest neighbor terms is increased. The $y=-1$ label can apparently
	only appear for a specific sign of the next-nearest neighbor phases.}
	\label{fig:haldane_distribution_biased}
\end{figure}

By measuring the frequency of the features, distinguished by class labels,
cf.~\figref{fig:haldane_distribution_biased}, we make an interesting
observation. Apparently, choosing the symmetry in the particular way that we
did, introduced a certain bias to our model. As a consequence, the
nearest-neighbor hopping term is now completely irrelevant for the classification. The
next-nearest neighbor terms, though, are showing improved contrast, since there
is less possibility for noise, which is also apparent in the redundancy and
joint PDF, cf.~\figref{fig:pearson_haldane_unbiased}. While we are able to detect a redundancy in \figref{fig:pearson_haldane_unbiased}(a), the values are still rather small. As a consequence we can regard the parameters as mostly independent and consider their marginal distributions.
The $C=-1$ phase was not produced in a statistically relevant sample size. We can relate this to the fact that we chose the correlations of the $C=+1$ phase when setting up symmetries and that the reference point is much closer to the $C=+1$ phase. Implementing the correlations between the nearest-neighbor hoppings via a sign change will result in a data set with a majority of samples belonging to the $C=-1$ class.

\section{General honeycomb lattice}
\label{sec:honeycomb}
\begin{figure*}
	\includegraphics[scale=1]{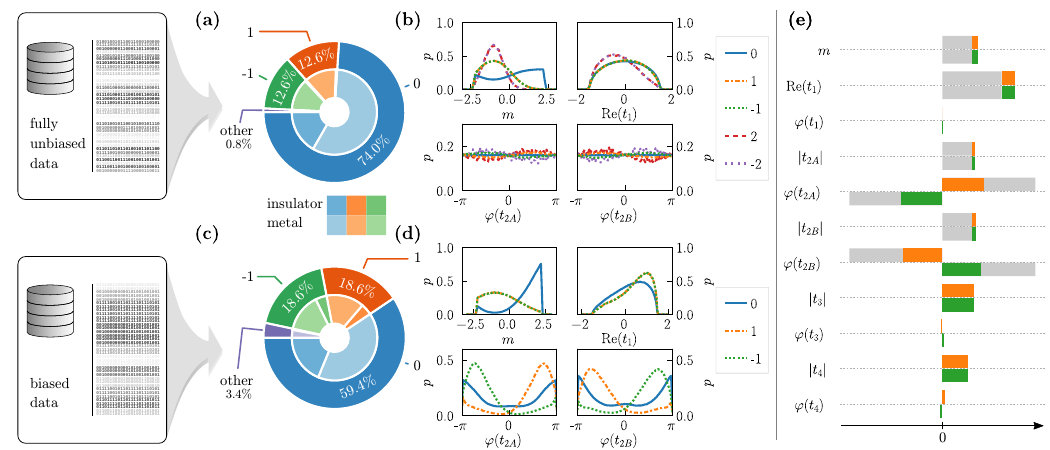}
	\caption{Results for the honeycomb lattice. Top row: fully unbiased model, bottom row: symmetrized model. \textbf{(a)} The fraction of topological samples is comparable to the Haldane case in the unbiased data, but much smaller in the symmetric data because the reference point is far away from a topological phase. While the majority of samples are metallic, all samples have separable bands. \textbf{(b)} PDFs of features with highest importance score $D_B$, separated from the rest by an order of magnitude. For the unbiased model we use all data points, while in the biased case we restrict to insulators only. \textbf{(c)} and \textbf{(d)} the same representation of the data for the biased calculation. The phases of the next-nearest neighbor hoppings are now a strong indicator for the topological phase. In \textbf{(e)} we show the overlap of the features with the parameters of a generic Haldane model (grey) for the $C=1$ (orange) and $C=-1$ (green) phase. The effective model contains the characteristics of the Haldane model.}
	\label{fig:honeycomb_results}
\end{figure*}

So far the reference point was carefully chosen to represent the Haldane model and located close to a non-trivial phase to make sure that both trivial and non-trivial samples are produced. In this section we want to test if our analysis also works for cases where no prior information is known. Therefore, we start from a very general honeycomb lattice, where we choose the reference point as
\begin{equation}
	x_\mathrm{ref} = (t_0^A, t_0^B, t_1^1, t_1^2, t_1^3, t_2^1, t_2^2, t_2^3, t_2^4, t_2^5, t_2^6, \ldots),
\end{equation}
where $t_i=1/d_i$ is chosen to be the inverse distance of the respective link. $t_0^A$, $t_0^B$ are set to 0 and 1, respectively, which fixes the scale and units of energy.
For the honeycomb lattice odd neighbors come in triplets and even neighbors come in sixtuplets. Therefore, we can write
\begin{equation}
	x_\mathrm{ref} = \left(0, 1, \frac{1}{d_1}, \frac{1}{d_1}, \frac{1}{d_1}, \frac{1}{d_2}, \frac{1}{d_2}, \frac{1}{d_2}, \frac{1}{d_2}, \frac{1}{d_2}, \frac{1}{d_2}, \ldots\right)
\end{equation}
with $d_1=1, d_2=\sqrt{2}, \ldots$.
This constitutes a rather generic but realistic \textit{a priori} ansatz that is known to be topologically trivial.
We run a fully unbiased sweep without assuming any symmetries and obtain the data presented in the top row of \figref{fig:honeycomb_results}.

Despite the presumably large distance of the reference point to a topologically non-trivial phase we obtain a reasonable number of non-trivial samples [\figref{fig:honeycomb_results}(a)].
Apparently, regardless of the greatly increased number of degrees of freedom, the phases of the hopping terms are revealed to be distinctly important, second only to the mass term. We take a look at the PDFs of these features in \figref{fig:honeycomb_results}(b) and observe that the phases for the next-nearest neighbor hoppings are split into two distinct categories.
We note that the sign of the class index is reflected in the distribution of
the next-nearest neighbor terms. In addition to the known phases from the
Haldane model we observe also larger indices $\pm 2$ and $\pm 3$ (not shown).
We compare the PDFs within the four different classes of hopping parameters in
terms of $D_B(p_{t_{1,i}}, p_{t_{1,j}})$ etc., and observe that all distributions
are very similar, except the ones of $t_{2A}$ and $t_{2B}$. This observation lends itself as an argument for introducing a symmetry between the hoppings with equal PDFs.

Taking into account this symmetry of the probability density functions and the correlations between features we reduce the model to a six-parameter model with $m, t_1, t_{2A}, t_{2B}, t_3, t_4$, which corresponds to 11 real features instead of the general 37.

Within this symmetrized (``biased") model (bottom row of \figref{fig:honeycomb_results}) we then observe a large number of different class labels. The $C=\pm 1$ classes that also appeared in the Haldane model represent by far the largest group of the non-trivial data and show very similar statistics, compare \figref{fig:honeycomb_results}(d) with \figref{fig:haldane_distribution_biased}. The phases of the next-nearest neighbor hoppings have a tendency towards opposite signs between $A$ and $B$ sublattices, which accounts for the vanishing net magnetic field. It is interesting how the added higher-order terms come into play. Statistically speaking, the added third and fourth nearest neighbor terms are irrelevant for the $C=\pm 1$ phase, which becomes apparent from the negligible deviation of their probability density functions from the base distribution and the absence of correlations. Obviously, samples of these two classes are continuously connected to the Haldane model. The new information here is that these phases are stable w.r.t.~noise and added longer range hopping terms.

During the sampling, especially in the general honeycomb model, it is clear that some combinations of parameters will not produce an insulating phase. Especially among the non-trivial data points we find only a small fraction to be insulating, while the majority lacks a band gap, cf.~\figref{fig:honeycomb_results}(a,c). However, in all cases we find topological bands that are clearly separable, which guarantees that the Chern index is well-defined. Although these phases are not insulators at all, we chose to keep them in the initial unbiased run to reduce the amount of samples needed. In fact, comparing the distributions between the topological metals and the topological insulators reveals that the key features are the same, i.e.~it is not strictly necessary to discard these data points, although the contrast, and therefore the amount of information, is higher for the insulating phases due to reduced noise. This is reflected in higher importance scores for all features in the topological insulator set compared to the topological metal set. It is possible to increase the insulating fraction by choosing the distribution observed for the topological insulator instead of the uniform distribution for the sampling process. This could be interpreted as learning the ideal distribution for generating topological insulators by looking at a completely unbiased data set, but performs less than ideal due to the assumption of independence during the sampling process.

In case the features are uncorrelated we can extract an effective model for each topological phase by looking at the peaks and average of the PDFs for each class label. More information, however, is encoded in the PDFs themselves and can be readily inspected due to the dimensional reduction. This information can be a guide to form a decision tree, i.e.~understand which parameters must be taken to produce a topological insulator.

The effective model found by our algorithm is shown in \figref{fig:honeycomb_results}(e). For both the unbiased and biased parameter selection we observe the characteristic features of the Haldane model with an added phase on the nearest neighbor hopping and real third- and fourth-neighbor hopping. The latter terms have already been found to be rather unimportant, i.e.~the occurrence in our effective model is entirely due to the reference point.
The beauty of this result is that by starting from a completely generic topologically trivial honeycomb model we reproduced the Haldane model as \emph{the} characteristic topological Chern insulator by purely statistical means. Although we did introduce a bias to combat the noise in the data there are traces of the Haldane model already visible in the unbiased data set. The effective models for the $C=+1$ and $C=-1$ phase differ only in the sign of the phase in the next-nearest neighbor hopping as is known from Haldane's original work \cite{Haldane1988}.

\section{Conclusion \& Outlook}
\label{sec:conclusion}
We have presented a scheme to learn the characteristics of topological phases and extract minimal models for a specific lattice. Using methods from  data science and statistics toolbox we performed dimensional reduction on an initially large feature space by extracting the most relevant features for the classification of each phase. Methods like these are essential to the construction of efficient machine learning models. We chose here to inspect only the statistical distributions of the individual parameters and their correlations between one another given a particular topological class, which comes at comparably low computational cost, and found that these quantities already contain enough information to extract a prototypical model for each topological phase.
In particular, by starting from a generic (far from topological) honeycomb model, we recovered the prototypical Haldane model as \emph{the} topological model in the Altland-Zirnbauer class A for the honeycomb lattice. It is expected that the method works even better for symmetry protected phases due the much lower potential for noise in models with fewer free parameters.
While the presented results are valid only for the non-interacting regime one can use a similar approach to learn about possible topological phases in interacting systems \cite{Mertz2019}.

Our method relies mainly on the inspection of integrated quantities, i.e.~distribution functions where all but one features are integrated out. This raises the question if this can still be useful since more often than not phase boundaries are complicated functions of many if not all parameters of the model. However, we have observed that our approach captures the exact same physics as e.g.~the permutation importance of random forests at much lower computational cost. In the present work correlations between pairs of features are taken into account, where we constrain the algorithm to features regarded as important in the first place.

The method presented here exploits the typical characteristics of phase diagrams, i.e.~that phases are not randomly distributed throughout the parameter space but follow particular patterns. As a consequence, not all parameter values will be equally likely to generate a particular phase, provided that the phase boundary crosses the parameter axis. This type of analysis works irrespective of the types of phases studied and it is not necessary to recognize the physical concepts underlying the different phases.

By using the bare tight-binding parameters as features we maximize the potential of learning comprehensible information about the data itself, since these parameters carry a straight-forward meaning. The success of the method shows that this information can be easily extracted.

Engineering new features in the data processing phase would allow for a more quantitative description of the phase diagram. To this end one could make use of higher-order correlation functions and try to maximize the importance score of a proposed new feature in an iterative learning algorithm. The prospects of such a method highly depend on the complexity of the model, though.

We note that this method is not in competition with neural network classifiers such as \cite{Zhang2017,Chng2017}, which attempt to learn the physics underlying the data. Although the importance of parameters w.r.t.~a particular classification can in principle be extracted from both methods, this process is much more difficult for sufficiently complex neural networks. In addition, we have shown that for this purpose training is not needed.

Besides finding topological models for arbitrary lattices, as demonstrated, the method can be applied for
data preparation and feature engineering for machine learning. In particular, by choosing fitted or ab-initio computed parameters as a
starting point our method can easily predict the possibility of engineering a topological phase for that particular material as well as a guide to how one could achieve this goal.
The task of extracting a prototypical model can be accomplished much easier than with a complicated machine learning model, which by construction is good at predicting but hard to understand. A way to combine both approaches would be to increase the interpretability of machine learning, which has been a highly active field of research in recent years \cite{Kim2016,Velez2017,Miller2017,Ghorbani2017}. By performing feature optimization to reduce the complexity of the model we have applied one possible step in this direction in the present work.

\begin{acknowledgments}
TM thanks Daniel Guterding for useful discussions. We thank Karim Zantout for reading the draft and his suggestions and acknowledge support by the Deutsche Forschungsgemeinschaft (DFG, German Research Foundation) through TR 288 - 422213477 (project B05).
\end{acknowledgments}

\bibliography{bibliography}

\end{document}